\documentclass{PoS}
\usepackage{verbatim}

\newcommand{\apjl}{ApJL}
\newcommand{\apj}{ApJ}

\newcommand{\aap}{A\&A}

\usepackage{xspace}
\usepackage{booktabs}

\newcommand{\fl}{{\it Fermi}-LAT\xspace}
\newcommand{\xrt}{{\it Swift}-XRT\xspace}
\newcommand{\hess}{H.E.S.S.\xspace}
\newcommand{\naima}{{\tt NAIMA}\xspace}

\title{Gamma-ray spectrometry in the cutoff region as a key for the understanding of radiation processes in 3C 279}

\ShortTitle{Gamma-ray spectrometry in the cutoff region}

\author{\speaker{Davit Zargaryan}\\
        Dublin Institute for Advanced Studies, 31 Fitzwilliam Place, Dublin 2, Ireland\\
        Russian-Armenian University, 123 Hovsep Emin str., 0051 Yerevan, Armenia\\
        E-mail: \email{dzargaryan@cp.dias.ie}}

\author{Carlo Romoli\\
        Max-Planck-Institut für Kernphysik P.O. Box 103980, 69029 Heidelberg, Germany\\
        E-mail: \email{Carlo.Romoli@mpi-hd.mpg.de}}
        
\author{Felix Aharonian\\
        Dublin Institute for Advanced Studies, 31 Fitzwilliam Place, Dublin 2, Ireland\\
        Max-Planck-Institut für Kernphysik P.O. Box 103980, 69029 Heidelberg, Germany\\
        E-mail: \email{Felix.Aharonian@mpi-hd.mpg.de}}

\abstract{We present results of spectrometric studies based on the observations of very strong 3C 279 flares in high and very-high-energy bands and
discuss their implications regarding the origin of radiation mechanisms. The FSRQ 3C 279 (z=0.536) is one of the most luminous gamma-ray
emitting AGN. It shows variability on time scales down to minutes during strong flares detected by the Fermi-LAT. We have analyzed LAT and
Swift-XRT data for the flaring periods in June 2015 and January 2018, when the flux above 100 MeV during hourly time intervals could exceed
$3 \times 10^{-5}$ ph cm$^{-2}$ s$^{-1}$. The X-ray flux derived from Swift X-ray observations described by a very hard spectrum with photon index ~1.5 is typically explained by
Inverse Compton scattering of low energy electrons and protons. Here we consider an alternative interpretation which describes the entire band
from X-ray to very high energy as a results of synchrotron emission by ultra relativistic electrons or protons.}

\FullConference{The High Energy Phenomena in Relativistic Outflows VII\\
                9-12 July, 2019\\
                Barcelona, Spain}

\begin{document}

\section{Introduction}
\label{sec:intro}

The term blazar indicates a subclass of Active Galactic Nuclei (AGN) and in this family we observe some of the most violent sources in the Universe. In the view of the {\it unified model} of AGNs, blazars are characterized by highly relativistic jets directed toward our line of sight \cite{begelman1984,urry1995}.
Based on the optical spectral emission lines, historically blazars are divided into two main subclasses: i) Flat Spectrum Radio Quasars (FSRQs), characterized by prominent emission lines ii) BL Lacs, with very weak or no spectral lines.

The broadband SED (Spectral Energy Distribution) of blazars is dominated by the non-thermal radiation from the jet, on top of the thermal radiation emitted by the Broad Line Region (BLR), accretion disk and torus. The non-thermal component of the SED exhibits a two-bump structure. Depending on the type of source, the first component extends from radio to X-ray energies, while the second component usually peaks at gamma ray energies, from hundreds of MeV to TeV energy.

Leptonic models generally explain the low energy component by synchrotron radiation from relativistic electrons accelerated in the jet. The high energy component would be instead the product of inverse Compton scattering of relativistic electrons on low-energy target photons from the same synchrotron radiation (SSC; Synchrotron Self-Compton, e.g. \cite{band1985,bloom1996}) and/or from external photon field (EC; External Compton) provided e.g. from the BLR (\cite{sikora1994}), a dusty torus (\cite{blazejovski200}) or the accretion disk.

On the other hand, an alternative explanation for origin of the high energy gamma-ray emission (second hump) is presented in terms of hadronic scenarios i.e. neutral pion decay, proton synchrotron. In particular, proton synchrotron emission has been invoked for interpretation of the TeV gamma-ray emission of blazars such as Mrk~501 \cite{2000NewA....5..377A}. In this case, the proton synchrotron scenario can explain also the fast variability of high energy gamma-rays, through a combination of large magnetic field and high Doppler boosting of the emission region. 

The spectrum of the accelerated particles responsible for the emission can be generally described with a power-law function with a modified exponential cut-off as in equation~\ref{eq:plsec}.
\begin{equation}
\frac{dN}{dE} = N_0\left(\frac{E}{E_0}\right)^{-\alpha}\exp\left[-\left(\frac{E}{E_c}\right)^{\beta}\right]
\label{eq:plsec}
\end{equation}
where $N_0$ is the flux normalization at the energy $E_0$, $\alpha$ is the power law index for the particle spectrum, $E_c$ is the cut-off energy and $\beta$ is the cut-off parameter that describes the steepness of the cut-off: sharp cut-offs have $\beta>1$, while for slow cut-offs $\beta<1$. The parameter $\beta$ value carries an imprint of the balance between acceleration and cooling processes. Furthermore, this functional form is generally maintained in the emitted photon spectrum and relations exist to estimate the cut-off parameter of the photons from the one of the particles, depending on the emission process \cite{2019arXiv190711663D,Kafexhiu2014,2012ApJ...753..176L}.

The blazar 3C~279 is a prominent gamma-ray emitting AGN located at $z=0.536$. It has been discovered in MeV/GeV gamma-rays by EGRET \cite{Hartman 1999}. A tentative very high energy (VHE; E > 100~GeV) signal has been reported by the MAGIC collaboration \cite{2008Sci...320.1752M}. Fermi-LAT has revealed the highly variable feature of the source regarding both the large amplitude of the flux variations and the short (several minute) variability timescales \cite{2016ApJ...824L..20A}.

Below we analyze the strong gamma-ray flares of 3C 279 in 2015 and 2018, and briefly discuss the origin of the gamma-ray emission.

\section{Data Analysis}
\label{sec:dataanalysis}

This work makes use of the high energy gamma-ray and X-ray data obtained by the \fl and \xrt instruments respectively. We reanalysed the data taken during the outbursts of 3C~279 in June 2015 and January 2018.

\subsection{Fermi-LAT} \label{subsec:latana}
The \fl is a pair conversion telescope designed to detect $\gamma$ rays with energies from 20 MeV to more than 1 TeV, measuring their arrival time, energy, and direction \cite{Atwood_2009}. We optimize the gamma-ray model of the Region of Interest (ROI) using the Fermi Science Tools version v11r5p3 and the {\tt fermipy} package adjusting the analysis for the latest P8R3 instrument response functions \cite{2013arXiv1303.3514A,2017ICRC...35..824W}. The data cover the period MJD 57187 to 57190 ( 14-17 Jun 2015, first flare) and MJD 58135.125 to 58137.125 (17-19 January 2018, second flare). In the energy range 0.1-300 GeV, we consider only the so-called SOURCE class events (evclass=128) and analyze a $12^{\circ} \times 12^{\circ}$ ROI centered on the position of 3C 279.
To avoid contamination of $\gamma$ rays originating from the Earth limb, an additional cut is set, allowing only photons with zenith angle $<90^{\circ}$. The Galactic and isotropic background components are included using the LAT standard diffuse background models gll\_iem\_v07 and iso\_P8R3\_SOURCE\_V2\_v1 respectively. During the analysis, the normalization of background models as well as the flux and spectral index of the 4FGL \cite{2019arXiv190210045T} sources within $10^{\circ}$ from 3C~279 are left as free parameters.

\subsection{Swift-XRT}
\label{subsec:swiftana}
The XRT data for the both flaring periods have been analyzed with XRTDAS
(v.3.3.0) using the standard procedure and the most recent calibration databases\footnote{https://www.swift.ac.uk/analysis/xrt/}. 
The observations were made in photon counting (PC) mode and for some observations the count rate was above 0.5 count/sec. This pileup effect was removed by excluding the events within a 4-pixel radius circle centered on the source position. Then, for the corrections of PSF losses the ancillary response files were generated through the {\tt xrtmkarf} command. 
Events for the spectral analysis were selected within a 20 pixel ($47''$) circle centered at the source position, while the background was extracted from an annulus with the same center and inner and outer radii of 51 ($120''$) and 85 pixels ($200''$), respectively.
The individual spectra at the energy range 0.3-10 keV are fitted with XSPEC v12.9.1a adopting an absorbed power-law model with a column density $N_{H} = 2.24 \times 10^{20}$ cm$^{-2}$.

\begin{table}[ht]
\footnotesize
\centering
\begin{tabular}{ c c c c c }
  \multicolumn{5}{c}{First Flaring Period -- 2015} \\
  \multicolumn{5}{c}{Swift-XRT} \\
  \hline
  Interval &Date &Index & $\nu F \nu \times$ 10$^{-11}$ (erg cm$^{-2}$ s$^{-1}$) &reduced ${\chi}^2$\\
  \hline
  interval 2 & Jun 15 & $1.26\pm0.03$ & $5.75\pm0.19$ & 0.87\\ 
  interval 4 & Jun 16 & $1.31\pm0.05$ & $10.47\pm0.17$ & 1.12\\
  interval 6 & Jun 17 & $1.32\pm0.09$ & $4.25\pm0.11$ & 0.79\\
  \hline
  \multicolumn{5}{c}{Fermi-LAT} \\
  \hline
  Interval &Date(MJD) &Index &Flux $\times 10^{-5}$(photon cm$^{-2}$ s$^{-1}$) &TS\\
  \hline
  Interval 1 & $57187.0-57188.25$ & $2.17\pm0.03$  & $1.56\pm0.06$ &4560.81\\
  Interval 2 & $57188.25-57188.625$ & $2.25\pm0.05$  & $2.23\pm0.01$ &2490.20\\
  Interval 3 & $57188.625-57189.0$ & $2.11\pm0.05$  & $1.17\pm0.07$ &1803.07\\
  Interval 4 & $57189.0-57189.25$ & $2.04\pm0.01$  & $3.34\pm0.10$ &12308.78\\
  Interval 5 & $57189.25-57189.625$ & $2.04\pm0.02$  & $2.66\pm0.06$ &16409.16\\
  Interval 6 & $57189.625-57190.0$ & $2.08\pm0.05$  & $1.02\pm0.06$ &1818.38\\
  \hline
  \multicolumn{5}{c}{Second Flaring Period -- 2018} \\
  \multicolumn{5}{c}{Swift-XRT} \\
  \hline
  Interval &Date &Index & $\nu F \nu \times$ 10$^{-11}$ (erg cm$^{-2}$ s$^{-1}$) &reduced ${\chi}^2$ \\
  \hline
  interval 1 & Jan 17 & $1.31\pm0.05$ & $8.36\pm0.19$ & 0.78\\
  interval 2 & Jan 18 & $1.27\pm0.05$ & $9.25\pm0.14$ & 0.92\\  
  interval 3 & Jan 19 & $1.34\pm0.08$ & $4.14\pm0.17$ & 0.89\\
  \multicolumn{5}{c}{Fermi-LAT} \\ 
  \hline
  Interval &Date-MJD &Index &Flux $\times 10^{-5}$(photon cm$^{-2}$ s$^{-1}$) &TS\\
  \hline
  Interval 1 & $58135.125-58136.125$ & $2.10\pm0.02$  & $1.83\pm0.05$ &8873.09\\
  Interval 2 & $58136.125-58136.5$ & $2.13\pm0.04$  & $2.47\pm0.13$ &2839.16\\
  Interval 3 & $58136.5-58137.125$ & $2.19\pm0.03$  & $1.95\pm0.03$ &5444.67\\
\end{tabular}
\caption{The parameters derived from the spectral analysis of \xrt and \fl data.}
\label{tab:fullphfits}
\end{table}

\subsection{Temporal analysis}
\label{tempana}

\begin{figure*}
\center
\includegraphics[width=0.48\textwidth]{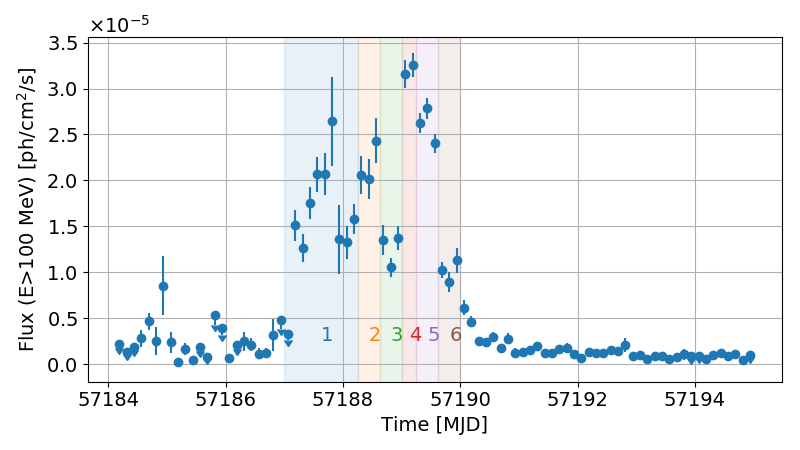}
\includegraphics[width=0.48\textwidth]{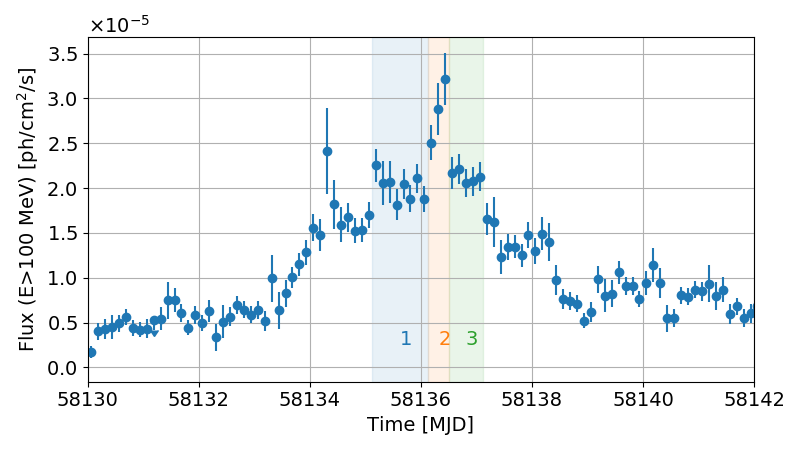}
\caption{The gamma-ray light curve during the flaring periods. In figures, six (for 2015 flare - left panel) and three (for 2018 flare - right panel) time intervals with the strongest gamma-ray fluxes are highlighted.\label{fig:lightcurves}}
\end{figure*}

\begin{figure}
    \centering
    \includegraphics[width=0.48\textwidth]{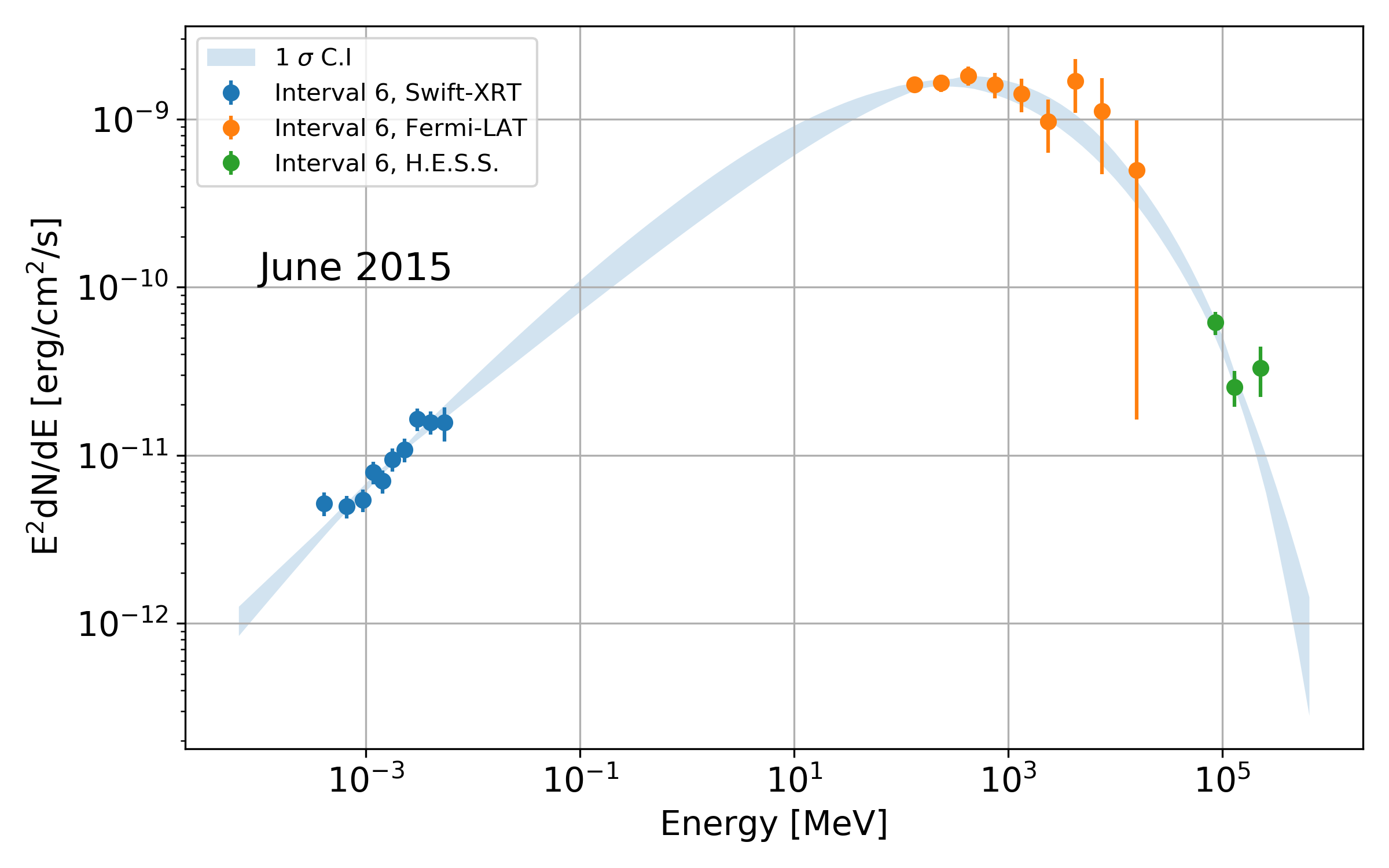}
    \includegraphics[width=0.48\textwidth]{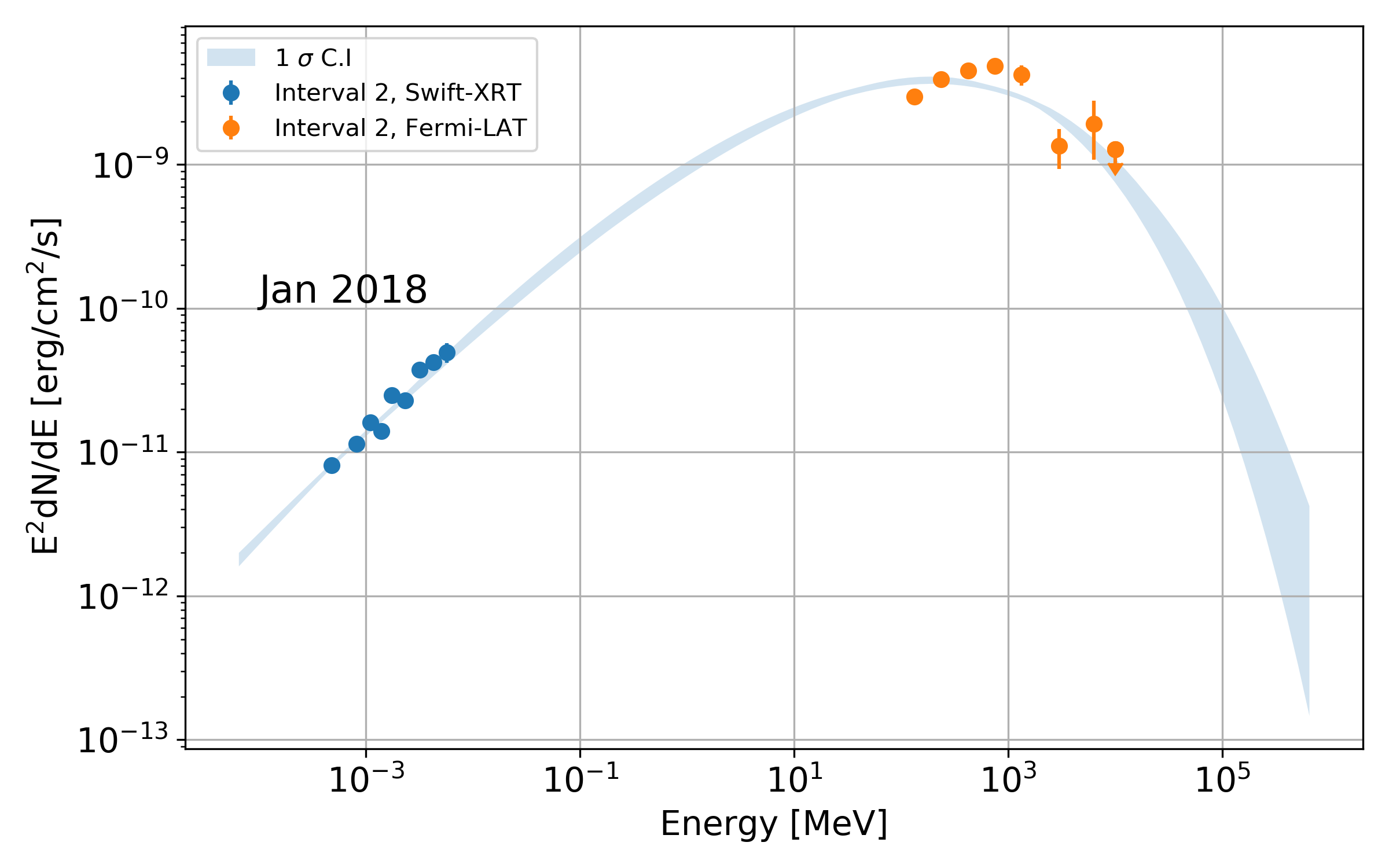}
    \caption{{\bf Left:} SED for the Interval 6 of the June 2015 flare. The blue points correspond to the Swift-XRT data, orange circles are from the \fl and the green ones are from \hess. The light-blue shaded area is the $1\sigma$ confidence interval of the \naima fit using proton synchrotron in turbulent magnetic field. {\bf Right:} Same as left panel but for interval 2 of the January 2018 flare.}
    \label{fig:photonfits}
\end{figure}

Here we focus on the study of the flaring states of 3C~279 in June 2015 and January 2018. The lightcurves were produced with a binning of 3 hours and different flux states to derive spectral points were defined through a Bayesian Blocks approach \cite{2013ApJ...764..167S}. For the implementation we used the {\tt astropy} PYTHON package\footnote{{\tt astropy} version 2.0.8} assuming a false alarm probability of 5\%. The result for the highest states of the flare are shown in Figure~\ref{fig:lightcurves}. Swift-XRT data are available for all 3 intervals of the January 2018 flare and for intervals 2, 4 and 6 of the June 2015 flare. For interval 6 of the June 2015 flare, H.E.S.S. data are also available. Here we used the points taken from \cite{2019A&A...627A.159H}.
The spectral analysis results for each time interval are summarized in Table~\ref{tab:fullphfits}, while the high energy SED are shown in Figure~\ref{fig:photonfits} for interval 6 of the June 2015 events and interval 2 of the January 2018 flare.

With this arrangement, we have quasi simultaneous observations over the high energy peak of the SED of the source.

\begin{table}
\footnotesize
    \centering
    \begin{tabular}{lc|cccc}
    \toprule
    \toprule
    {\bf Interval} & & {\bf Total energy} [erg] &\textbf{Index} & \textbf{Cut-off} [$\log_{10}(E_c/1 \rm{ TeV})$] & \textbf{Cut-off Index} \\
    \midrule
    \midrule
    Int. 2 2015 & protons& $(4.9\pm0.2)\times10^{48}$ & $1.4\pm0.1$ & $3.9^{+0.3}_{-0.4}$ & $0.55\pm0.10$\\
     & electrons          & $(3.3\pm0.3)\times10^{43}$ & $1.4\pm0.1$ & $0.0\pm0.4$ & $0.56\pm0.10$ \\
     \midrule
    Int. 4 2015 & protons & $(7.3\pm0.3)\times10^{48}$ & $0.9\pm0.2$ & $1.3\pm1.0$ & $0.25^{+0.05}_{-0.03}$\\
     & electrons          & $(5.0\pm0.2)\times10^{43}$ & $0.8\pm0.2$ & $-3.2\pm1.0$ & $0.23\pm0.04$ \\
     \midrule
    Int. 6 2015 & protons& $(2.9^{+0.5}_{-0.2})\times10^{48}$ & $1.78^{+0.12}_{-0.15}$ & $4.4^{+0.4}_{-0.5}$ & $0.62^{+0.15}_{-0.11}$\\
     & electrons          & $(1.7\pm0.1)\times10^{43}$ & $1.73\pm0.14$ & $0.4^{+0.4}_{-0.5}$ & $0.58^{+0.13}_{-0.10}$ \\
    \midrule
    \midrule
    Int. 1 2018 & protons &$(5.3\pm0.2)\times10^{48}$ & $1.5\pm0.2$ & $2.9\pm0.9$ & $0.34^{+0.09}_{-0.06}$ \\
     & electrons          & $(3.5\pm0.1)\times10^{43}$ & $1.5\pm0.2$ & $-0.8\pm0.8$ & $0.36^{+0.09}_{-0.06}$ \\
     \midrule
    Int. 2 2018 & protons & $(6.3\pm0.3)\times10^{48}$ & $1.5\pm0.2$ & $3.6\pm0.8$ & $0.44^{+0.17}_{-0.10}$\\
     & electrons          & $(4.2\pm0.3)\times10^{43}$ & $1.4\pm0.2$ & $-0.6^{+0.7}_{-0.9}$ & $0.4\pm0.1$ \\
     \midrule
    Int. 3 2018 & protons & $(3.6\pm0.2)\times10^{48}$ & $1.67^{+0.13}_{-0.2}$ & $4.6^{+0.3}_{-0.6}$ & $0.8^{+0.4}_{-0.3}$\\
     & electrons          & $(2.3\pm0.2)\times10^{43}$ & $1.66^{+0.14}_{-0.2}$ & $0.7^{+0.3}_{-0.6}$ & $0.8^{+0.5}_{-0.3}$ \\
    \bottomrule
    \bottomrule
    \end{tabular}
    \caption{Full results of \naima fits. Reference value for magnetic field is 100 G and 1 G (for protons and electrons respectively). Lorentz factor used is $\Gamma=25$ and a Doppler factor $\delta=50$. All values refer to the model with synchrotron emission in turbulent field. The kernel of the synchrotron emission is the same for protons and electrons. The differences in the spectral index and cut-off index values are only due to the statistical differences between the results of the 2 MCMC fits. }
    \label{tab:fullres}
\end{table}

\section{Theoretical Modelling}
\label{sec:modelling}

The SED points were de-boosted by the jet Lorentz factor $\Gamma$ and then fitted using a modified version of the \naima python package (version 0.8.3) \cite{naima}, where we implemented the synchrotron emission for protons and the synchrotron emission in turbulent magnetic fields for both electrons and protons as described in \cite{2019arXiv190711663D}. The fits were performed using 256 parallel walkers for 1500 steps with a burn-in phase of 300 steps (see \cite{naima} further explanations on this fitting technique).
A characteristic of the model of synchrotron emission in turbulent field is that it would require a slightly steeper cut-off to describe the same photon spectrum. The asymptotic relation between the cut-off parameter of the particle $\beta_p$ and the cut-off parameter of the photon $\beta_{\gamma}$ is \cite{2019arXiv190711663D}: $\beta_{\gamma} = \frac{2\beta_p}{3\beta_p+4}$ and not $\beta_{\gamma} =\frac{\beta_p}{\beta_p+2}$ as the theory for simple synchrotron emission predicts \cite{1989A&A...214...14F,2007A&A...465..695Z}.

\begin{figure}
    \centering
    \includegraphics[width=0.65\textwidth]{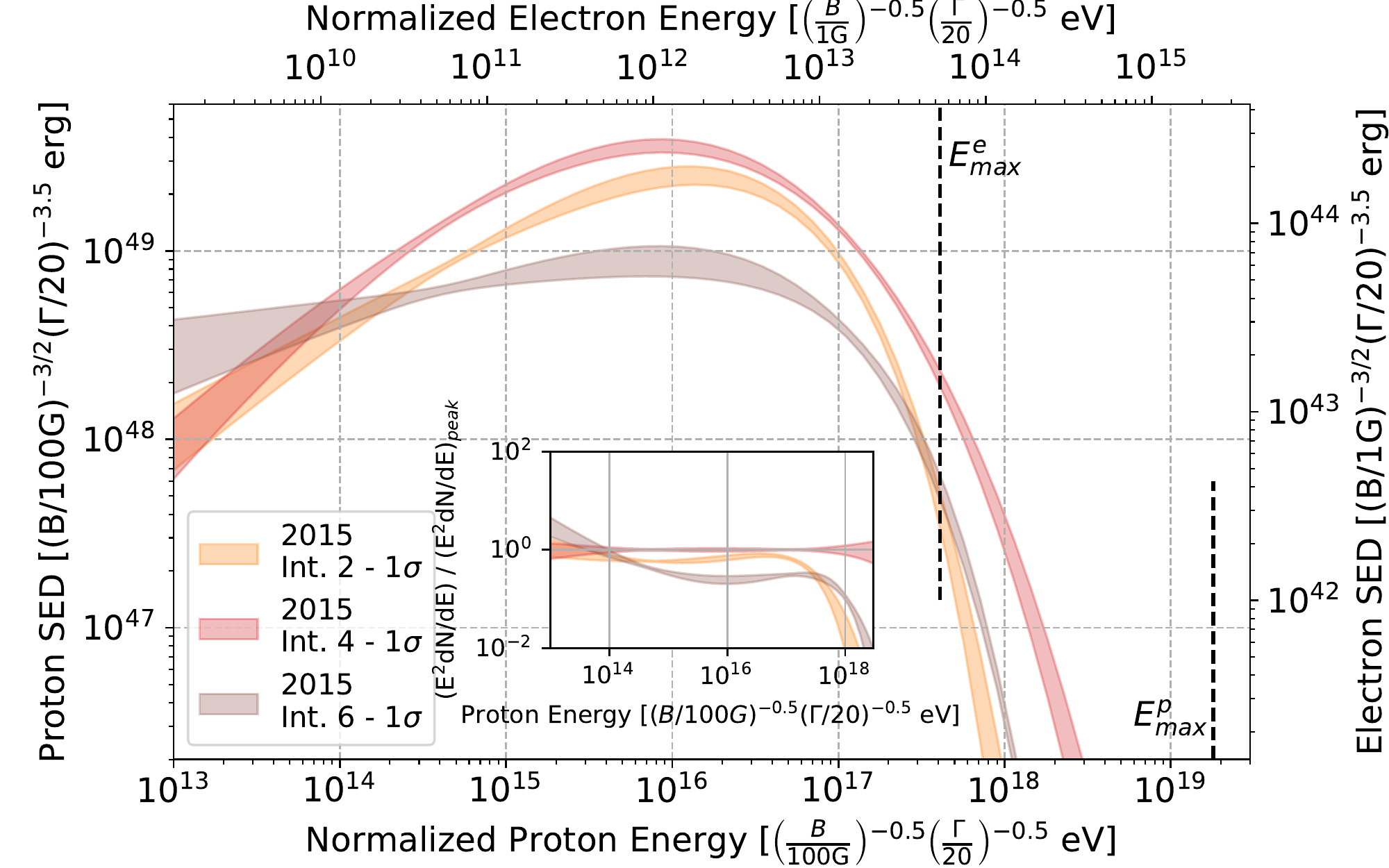}
    \caption{Spectral energy distribution for the primary protons(the lower X-axis) and electrons(the top X-axis) assuming that their synchrotron radiation is responsible for the radiation of three periods of the June 2015 flare. The bands represents the $1\sigma$ best fit obtained with \naima. The X-axes are scaled for the magnetic field and the Lorentz factor. The vertical dashed lines indicate the maximum energy for synchrotron emitting electrons and protons derived from the balance between the acceleration and energy loss rates. For the turbulent magnetic field, Gaussian distribution is assumed. For the acceleration rate, we assume $\eta=1$. The inset shows the ratio between the particle SED for the interval with maximum flux and the others. }
    \label{fig:particlesed15}
\end{figure}

\begin{figure}
    \centering
    \includegraphics[width=0.65\textwidth]{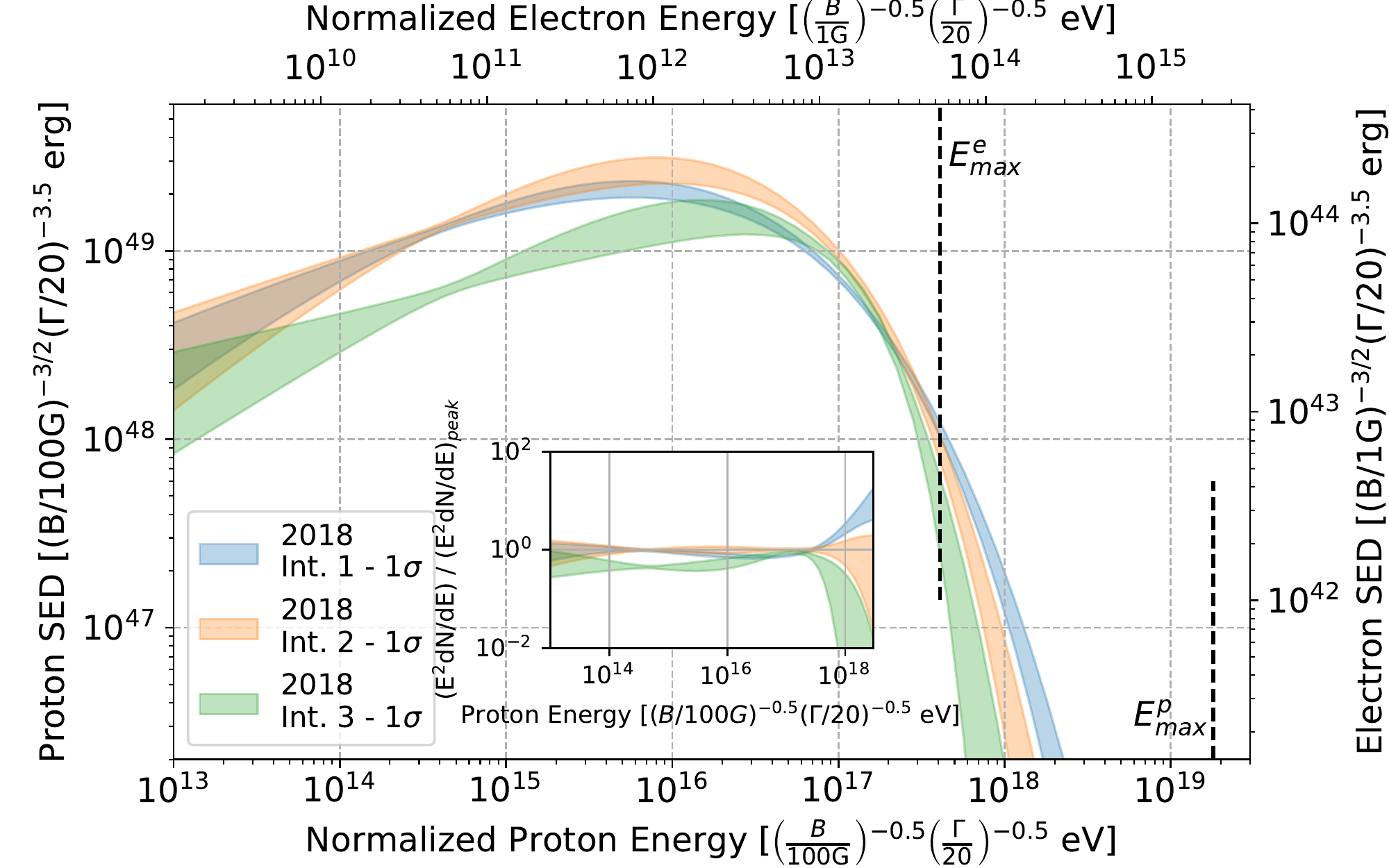}
    \caption{Same plot as in Figure~\ref{fig:particlesed15}, but for the flare of January 2018.}
    \label{fig:particlesed18}
\end{figure}

The SED points were fitted together to derive the primary particle spectrum in the assumption that the Doppler boosting is given by $\delta = 2\Gamma$.
The complete fit results are reported in Table~\ref{tab:fullres}.\textbf{}
Figures~\ref{fig:particlesed15} and \ref{fig:particlesed18}, show visually the resulting particle SED for electrons and protons that best fit the X- and $\gamma$-ray data using a magnetic field of 100 G for protons and 1 G for electrons. To make the plots more general, the axis show also the scaling relations for different choices of the magnetic field and the Lorentz factor.



\section{Discussion and Conclusions}
\label{sec:discussion}

Two exceptionally strong gamma-ray flares of 3C 279 in 2015 and 2018 have been accompanied by hard-spectrum non-thermal X-ray emission with photon index $\sim1.5$. Generally, non-thermal X-ray emission with very hard spectra (photon index $< 2$) from flaring AGN is interpreted as the continuation of the Inverse Compton radiation (responsible for high energy gamma-rays) to the X-ray domain.
In this work, we have explored an alternative interpretation that the entire region over 9 decades from X-rays to very high energy gamma-rays is described by a single mechanism - synchrotron radiation of electrons or protons.
Independently from the details, the interpretation within both proton-synchrotron and electron-synchrotron models, implies that we deal with an extreme accelerator boosting the particles to the maximum possible energies allowed by classical electrodynamics.

The major problem for the electron synchrotron scenario is the continuation of the particle distribution beyond the maximum energy set by the synchrotron losses, even assuming the maximum acceleration rate of electrons with $\eta=1$. This happens for reasonable Lorentz factor such as $\Gamma \sim20$.

For the synchrotron radiation of protons, the maximum energy requirement is substantially relaxed. Under the assumptions of figures~\ref{fig:particlesed15} and \ref{fig:particlesed18}, the data can be well described with proton distributions extending up to $10^{18}$ eV with stretched cut-offs and a total energy of the order of $10^{49}$ erg. The issues with this scenario however arise from the very large energetics.

\section*{Acknowledgement}
DZ acknowledges funding from the Irish Research Council Starting Laureate Award (IRCLA/2017/83)

\end{document}